\documentstyle[11pt,paspconf]{article}
\def\edcomment#1{\iffalse\marginpar{\raggedright\sl#1\/}\else\relax\fi}
\reversemarginpar

\begin{document}
\title{The Presence of Accretion Disks in Novae Shortly After their Outbursts}
\author{Alon Retter}
\author{Elia Leibowitz}
\affil{School of Physics and Astronomy and the Wise Observatory,
Raymond and Beverly Sackler Faculty of Exact Sciences, Tel-Aviv
University, Tel Aviv, 69978, Israel}

\begin{abstract}
It was believed, with little theoretical basis, that the accretion disc
(AD) is destroyed in nova outbursts, and recovers only a few decades
later.  We looked for observational evidence for the presence of ADs in
young novae. We discuss two cases: 1. Nova V1974 Cyg 1992 - we found
permanent superhumps in its light curve - a very strong evidence for
the presence of the AD according to the disc-instability model. 2. Nova
V1425 Aql 1995 - its possible classification as an Intermediate Polar
system suggests that it's most likely that the accretion is maintained
through an AD.
\end{abstract}

\section{Introduction}

It is not known what is the fate of the AD in a classical nova system
immediately following the outburst event. It was assumed, that it is
being destroyed by this cataclysmic eruption.  In addition, there are
no theoretical calculations concerning the question when is the AD
rebuilt in the remnant system.  Leibowitz et al. (1992) discovered an
eclipse three weeks after maximum light in Nova V838 Herculis 1991.
They interpreted it as the occultation of the AD by the secondary
star.  A major aim of A.R. Ph.D. thesis was to look for further
evidence for the presence of ADs in young (months-years old) novae.

\section{Observational results}

We describe the photometric results of two objects:

\subsection{Nova V1974 Cygni 1992}

Two distinct periodicities in the light curve of V1974~Cyg were
independently discovered by Semeniuk et al. (1995) and by Retter, Ofek
\& Leibowitz (1995). Semeniuk et al. suggested, that the 2.04 hr
period, which is larger than the second period by about 5\%, is the
spin period of the rotating white dwarf, and predicted that it will
continue its decrease towards the shorter assumed orbital period.
Retter et al. suggested a connection between the second periodicity of
the nova and the superhump phenomenon in the SU UMa stars, based on the
fact that the two periods of V1974 Cyg fits well within the Stolz \&
Schoembs (1984) relation for the two periods of SU UMa systems.

Retter, Leibowitz \& Ofek (1997) and Skillman et al. (1997) showed that
the longer period stopped the trend of decrease in 1995, and began to
increase during that year. They also listed many photometric features
in the light curve of the nova, that resemble the properties of
systems that are in a state of permanent superhumps.  They, therefore,
concluded that V1974 Cyg is also exhibiting the permanent superhumps
phenomenon.  Superhumps characterize the SU UMa class of CVs that are
known to have an AD in their underlying stellar system.  Thus the
observations in V1974 Cyg indicate the presence of an AD in that system
30 months after its eruption.

\subsection{Nova V1425 Aquilae 1995}

Retter, Leibowitz \& Kovo-Kariti (1997) found three periodicities in
the power spectrum of Nova Aql 1995. They interpret them as the orbital
period of the binary system, the spin period of a magnetic white dwarf
and the beat period between them. This suggests that the system belongs
to the Intermediate Polar group. Only one object out of the 13
Intermediate Polars listed by Hellier, 1996  (see his Fig. 2) is
believed to be a disc-less system. Based on this statistics, we may
regard it as very likely that no later than 1996 May, 15 months after
its outburst, Nova Aql 95 already possessed an AD within its binary
system.

\section{Summary}

Our observations on these two novae support the notion, that ADs do
exist in young novae, already a few months after the outburst.

\end{document}